\documentclass[pra,twocolumn,tightenlines,eqsecnum,showpacs,nofootinbib,floatfix]{revtex4}
\usepackage{pslatex}
\usepackage{amsmath}
\usepackage{epsfig}
\numberwithin{equation}{section}
\renewcommand{\theequation}{\arabic{section}.\arabic{equation}}

\begin{document}

\title{The Physics of `Now'}

\author{James B.~Hartle}

\email{hartle@physics.ucsb.edu}

\affiliation{Department of Physics\\
University of California, \\ Santa Barbara, CA 93106-9530}

\date{\today}

\vskip .2in

%\centerline{\Huge DRAFT \today}
\vskip .2in 
\begin{abstract}

The world is four-dimensional according to fundamental physics, governed
by basic laws that operate in a spacetime that has no unique division 
into space and time. Yet our
subjective experience of this world is divided into present, past, and
future. What is the origin of this division? What is its four-dimensional
description? Is this the only way experience can be organized consistently
with the basic laws of physics? This paper reviews such questions through
simple models of information gathering and utilizing systems (IGUSes) such
as ourselves.

Past, present, and future are not properties of four-dimensional spacetime
but notions describing how individual IGUSes process information. Their
origin is to be found in how these IGUSes evolved or were constructed. The
past, present, and future of an IGUS is consistent with the
four-dimensional laws of physics and can be described in four-dimensional
terms. The present, for instance, is not a moment of time in the sense of a
spacelike surface in spacetime. Rather there is a localized notion of
present at each point along an IGUS' world line. The common present of many
localized IGUSes is an approximate notion appropriate when they are
sufficiently close to each other and have relative velocities much less
than that of light.

Some features of the present, past, and future organization are closely 
related to basic physical laws. For example, the retarded nature of 
electromagnetic radiation and the second law of thermodynamics are the 
likely reasons we 
cannot remember the future.  But modes of organization that are 
different from present, past and future can be imagined 
that are also consistent with the basic laws
of physics. We speculate why the present, past, and future organization might
be favored by evolution and suggest that therefore it may be a cognitive
universal.

\end{abstract}
%\pacs{PACS }

\maketitle

\section{Introduction}
\label{intro}

A lesson of the physics of the last century is that, up here, on length
scales much greater than the Planck length, the world is four-dimensional 
with a
classical spacetime geometry. There is neither a unique notion of space nor a
unique notion of time. Rather, at each point in spacetime there are a
family of timelike directions and three times as many spacelike directions.
Yet, in this four-dimensional world, we divide our subjective experience up
into past, present, and future. These seem very different. We experience
the present, remember the
past, and predict the future. How is our experience
organized in this way? Why is it so organized? What is the four-dimensional
description of our past, present, and future? Is this the only way
experience can be organized? This paper is concerned with such questions.

The general laws of physics by themselves provide no answers.  
Past, present, and future are not properties of four-dimensional
spacetime. Rather, they are properties of a specific class of subsystems of
the universe that can usefully be called {\it information gathering
and utilizing systems} (IGUSes) \cite{GM94}. The term is broad enough 
to include both single representatives of biological species
that have evolved naturally and mechanical robots that  
were constructed artificially. It includes human beings both individually
and collectively as members of groups, cultures, and civilizations. 

To understand
past, present, and future it is necessary to understand how an IGUS employs
such notions in the processing of information. To understand why it is
organized in this way it is necessary to understand how it is constructed
and ultimately how it evolved. Questions about past, present, and future
therefore are most naturally the province of psychology, artificial
intelligence, evolutionary biology, and philosophy\footnote{See, e.g
\cite{But99} for a collection of philosophical papers on time. This paper
does not aim to discuss or resolve any of the philosophical debates
on the nature of time.}. 

However, questions concerning past, present, and future cannot be completely
divorced from physics. For instance, the notions must be describable in
four-dimensional terms just to be consistent with the fundamental picture
of spacetime. Further, as we will review, the distinctions between the 
past, present, and future of an IGUS depend 
upon  some of the arrows of time that our universe exhibits 
such as
that summarized by the second law of thermodynamics. In the 
tradition of theoretical physics, we illustrate these connections
with simple models of an IGUS --- achieving clarity at the risk of
irrelevance. Our considerations are entirely based on classical
physics.\footnote{Arrows of time in the context of quantum mechanics, as well
as the quantum mechanical arrow of time, are discussed in \cite{GH93b} in the
framework of a time-neutral generalized formulation of quantum theory. 
The  author knows of no obstacle of principle to extending the 
present classical discussion to quantum mechanics in that framework. For the 
special features of history in quantum mechanics see, {\it e.g.} \cite{Har98b}.} 
One such model IGUS --- a robot --- is described in Section \ref{II}. It is
simple enough to be easily analyzed, but complex enough to suggest how
realistic IGUSes distinguish between past, present, and future. The
four-dimensional description of this robot is discussed in Section
\ref{III}.
There we will see that the
robot's present is not a moment in spacetime. Rather, there is a present at each
instant along the robot's world line consisting of its most
recently acquired data about its external environment. The approximate
common notion of
`now' that could be utilized  by a collection of nearby robots moving
slowly with respect to one another is also described.   

Section \ref{IV} describes the connection of present, past and future with the 
thermodynamic arrow of time
and the radiation arrow of time. It seeks to answer the question,
``Could  we
construct a robot that would remember the future?" Section \ref{V} describes
alternative organizations of a robot's experience that are different from
past, present, and future. These are consistent with the four-dimensional
laws of physics. But the possibility of these alternative organizations shows that 
past, present, and future are not
consequences of these laws. We speculate, however, that, as a
consequence of the laws of physics,  the past, present,
and future organization may offer an evolutionary advantage over the other
modes of organization. This supports a conjecture that past, present, and
future may be a cognitive universal \cite{She94} for sufficiently localized 
IGUSes.

\section{A Model IGUS}
\label{II}

Imagine constructing a robot which gathers and utilizes information in the
following manner (Figure 1):

\begin{figure}[t]
%\centerline{\epsfig{file=nowfig1.ps, width=3.25in
%bblly=507, bburx=435, bbury=624, clip=true}}
\centerline{\epsfig{file=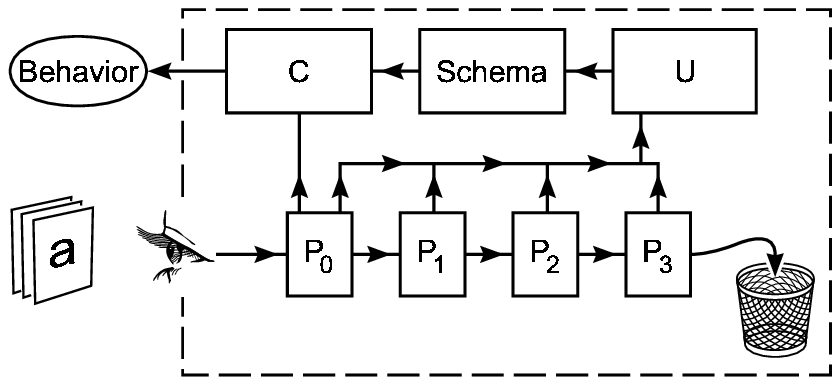, width=3.25in}}
\caption{Information processing in a model robot.  The information flow in
the robot described in the text is represented schematically in this 
diagram. The internal workings of the robot are within the dotted box; its 
external environment is without.  Every proper time interval $\tau_*$ the 
robot captures an image of its external environment. In the example 
illustrated, this is of a stack of cards labeled $a, b, c$, etc. whose top 
member changes from time to time.  The captured image is stored in register 
$P_0$ which constitutes the robot's present.  Just before the next capture 
the image is $P_3$ is erased and images in $P_0$, $P_1$, and $P_3$ are 
shifted to the right making room for the new image in $P_0$. The registers 
$P_1$, $P_2$, and $P_3$ therefore constitute the robot's memory of the past.
At each capture, the robot forgets the image in register $P_3$.
The robot uses the images in $P_0, P_1, P_2$, and $P_3$ in two processes 
of computation: $C$ (``conscious'') and $U$ (``unconscious''). The process 
$U$ uses the data in all registers to update a simplified model or schema 
of the external environment. That is used by $C$ together with the 
most recently acquired data in $P_0$ to make predictions about its 
environment to the future of the data in $P_0$, make decisions, and direct 
behavior accordingly.  The robot may therefore be said to experience 
(through $C$) the present in $P_0$, predict the future, and remember the
past in $P_1, P_2$, and $P_3$.}
\end{figure}

\vskip .10in
\noindent
{\sl Information Gathering:} The robot has $n+1$ memory locations
$P_0, P_1, \cdots, P_n$ which we call `registers' for short. These
contain a time series of images of its external environment assembled as
follows: At times separated by intervals $\tau_*$ the image in register
$P_n$ is erased and replaced by the image in $P_{n-1}$. Then the image in
$P_{n-1}$ is erased and replaced by the image in $P_{n-2}$, and so on. For
the last step, the robot captures a new image of its external environment
and stores it in register $P_0$. Thus, at any one time, the robot possesses
a coarse-grained image history of its environment extending over a time
$(n+1)\tau_*$. The most recent image is in $P_0$; the oldest is in $P_n$.
\vskip .10in
\noindent
{\sl Information Utilization:} The robot employs the information in
the registers $P_0, P_1, \cdots, P_n$ to compute predictions about its
environment at times to the future of the data in $P_0$ and direct its
behavior based upon these predictions. It does this in two steps employing
two different processes of computation:

\begin{itemize}
\item{\underline{Schema}:}  
The robot's memory stores a simplified model of its
environment containing, not all the information in $P_0, P_1, \cdots, P_n$,
but only those parts important for the robot's  functioning. This model is
called a {\it schema} \cite{GM94}. Each time interval $\tau_*$, the robot 
updates its
schema making use of the information in $P_0$ by a process of computation
we denote by $U$.

The schema might contain the locations and
trajectories of food, predators, obstacles to locomotion, fellow robots,
etc. It might contain hard-wired rules for success ({\it e.g.} get food
--- yes, be food --- no) and perhaps even crude approximations to the rules
of geometry and the laws of physics ({\it e.g.} objects generally fall
down).  It might contain summaries of regularities of the environment
abstracted from the information gathered long before the period covered by
registers $P_0, \cdots, P_n$ or explained to it by other robots,
etc.\footnote{A history book is a familiar part of the schema of the
collective IGUS linked by human culture. It is a summary and analysis of
records gathered at diverse times and places. That is true whether the
history is of human actions or the scientific history of the universe. The
schema resulting from the reconstruction of present records simplifies the
process of future prediction. (For more on utility of history, see
\cite{Har98b}.)} 
%Indeed, in human affairs ``those who cannot remember the
%past are condemned to repeat it'' \cite{San1905}.}

\item{\underline{Decisions and Behavior}:} Each time interval $\tau_*$ 
the robot
uses its schema and the fresh image in $P_0$ to assess its situation,
predict the future, and make decisions on what behavior to exhibit next by
a process of computation that we denote by $C$. This is distinct from $U$. The
important point for this paper is that the robot directly employs only the
most recently acquired image in register $P_0$ in this process of
computation $C$. The information in $P_1, \cdots, P_n$ is employed only through
the schema.
\end{itemize}

It seems possible that such a robot could be constructed.  As a model of
sophisticated IGUSes such as ourselves it is grossly oversimplified. Yet,
it possesses a number of features that are similar to those in sophisticated
IGUSes that are relevant for understanding past, present, and future:
The robot has a coarse-grained memory of its external environment contained
in registers $P_0, P_1, \cdots, P_n$. The robot has two processes of
computation, $C$ and $U$. Without entering into the treacherous
issue of whether the robot is conscious,
the two processes have a number of similarities with our own processes of
conscious and unconscious computation:

\begin{itemize}

\item $U$ computation provides input to decision-making $C$ computation.
 
\item There is direct input to $C$ computation only from the most recently
acquired image in the register $P_0$. The images in $P_1, \cdots, P_n$
affect $C$ only through the schema computed in $U$.

\item Information flows into and out from the register $P_0$ directly used
by $C$.

\end{itemize}

Equally evident are some significant differences between the robot and
ourselves.  Our information about the external environment is not exclusively 
visual, it is not stored in a
linear array of registers, nor is it transferred from one to the other in
the simple manner described. Input and records are not separated by sharp
time divisions. We can consciously access memories of other
than the most recent data, although often imperfectly and after modification
by unconscious computation. This list of differences can easily be
extended, but that should not obscure the similarities discussed above.

The analogies between the robot and ourselves can be emphasized by
employing everyday subjective terminology to describe the robot. For example, we
will call $C$ and $U$ computation `conscious' and `unconscious' secure in
the confidence that such terms can be eliminated in favor of the mechanical
description we have employed up to this point if necessary for clarity.
Proceeding in this way, we can say that the robot `observes' its
environment. The register $P_0$ contains a record of the `present', and
the registers $P_1, \cdots P_n$ are records of the
`past'\footnote{For the moment we take the records $P_1, \cdots
P_n$ to {\it define} the past. Section \ref{IV} will connect this notion
of past with other physical notions of `past', that defined by 
the the time direction toward  the big bang for instance.}. When the
register $P_n$ is erased, the robot has `forgotten' its contents. The
present extends\footnote{As James \cite{Jam1890} put it more 
eloquently: ``... the practically cognized present is no knife-edge,
but a saddleback, with a certain breadth of its own, on which we sit
perched, and from which we look in two directions in time.''}
over a finite interval\footnote{For human IGUSes the time $\tau_*$ can 
be taken to be of order the $.1$s separation time needed discriminate
between two visual signals \cite{Wessum}.} $\tau_*$.

The robot has conscious focus on the present, but only access to the
past through the records that are inputs to the unconscious computation 
of its schema.  The robot can thus be said to `experience' the present 
and `remember' the past.  The `flow of time' is the movement of 
information into the register of conscious focus and out again.
Prediction requires computation --- either conscious or unconscious
--- from memories of the present and past acquired by observation
and is thus distinct from remembering.

The subjective past, present, and future, the flow of 
time, and the distinction between predicting and remembering  are
represented concretely and physically in the structure and function
of the model robot.  We now proceed to describe this structure and
function in four-dimensional terms.

\section{The Present is Not a Moment in Time}
\label{III}

This section gives the four-dimensional, spacetime description of the robot
specified in the previous section.\footnote{An abbreviated version of this
discussion was given in \cite{Har98b}.}   For simplicity we consider the flat
spacetime of special relativity (Minkowski space). But with little change
it could be a curved spacetime of general relativity.

\subsection{Some Features of Minkowski Space}

\begin{figure}[t]
\centerline{\epsfig{file=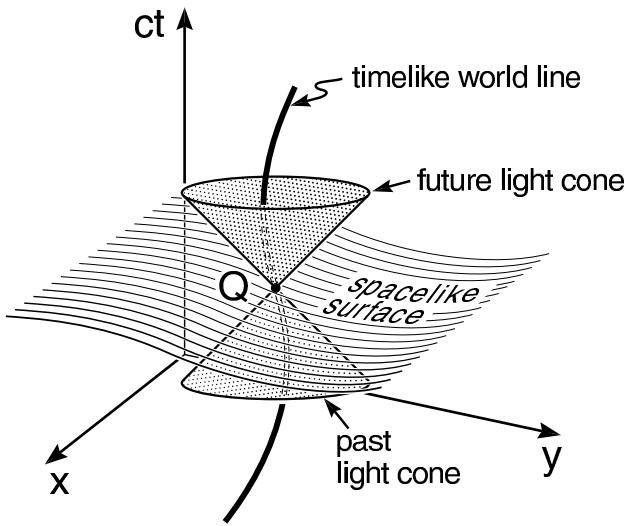, width=3.25in}}
\caption{Spacetime Concepts. This spacetime diagram represents a
three-dimensional slice of a four-dimensional flat spacetime defined by
three axes $ct$, $x$, and $y$ of the four specifying an  inertial (Lorentz)
frame.  The ideas of event, light
cone, world line, and spacelike surface are illustrated. 
An event is a point in spacetime like $Q$. Each point has a future
and past light cone.  
A spacelike surface like the one illustrated defines an instant in time. 
Each such surface divides
spacetime into two regions conventionally called the future of this surface
and its past. There are an infinity of such families and thus infinitely
many different ways of defining instants in time and their futures and
pasts.  In the context of cosmology,
the past of a spacelike surface is defined to be the region closest to the
big bang and the future is the region furthest away.   
}
\end{figure}

We begin by recalling\footnote{For a classic text on special relativity
from a spacetime point of view, see \cite{TW63}}
a few important features of four-dimensional spacetime that are illustrated in
Figure 2. Events occur at points in spacetime. At each point $Q$ there is a
{\it light cone} consisting of two parts. The future light cone is the
three-dimensional surface generated by the light rays emerging from $Q$. The
past light cone is similarly defined by the light rays converging on $Q$.
(The labels `future' and `past' are conventions at this point in the
discussion. In Section \ref{IV}  we will define them in the cosmological context.)  Points inside the light cone of $Q$ are timelike separated from
it; points outside the light cone are spacelike separated. Points inside
the future light cone of $Q$ are in its future; points inside the past light
cone are in its past. Points outside the light cone are neither.

The center of mass of a localized IGUS, such as the robot of the previous
section, describes a timelike world line in spacetime. At each point along
the world line, any tangent to it lies inside the light cone so that the
IGUS is moving at less than the speed of light in any inertial frame.

A moment in time is a three-dimensional spacelike surface in spacetime --- one in which any two nearby
points are spacelike separated. Each spacelike surface divides spacetime
into two regions --- one to its future and one to its past. A family of
spacelike surfaces such that each point in spacetime lies on one and only
member of the family  specifies a division of spacetime into space and time.
The family of surfaces defined by constant values of the time of a
particular  inertial frame is an example (e.g. the surfaces of constant
$ct$ in Figures 2-4). 
Different families of spacelike surfaces define different notions of space
and different notions of time none of which is preferred over the other.

\subsection{The Past, Present, and Future of the Robot}

\begin{figure}[t!]
\centerline{\epsfig{file=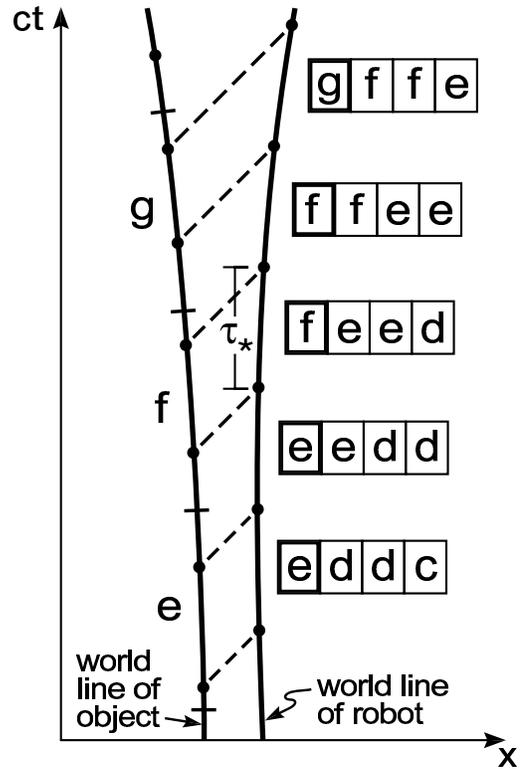,height=4.00in}}
\caption{A spacetime description of the present and past of the robot whose
information processing is illustrated in Figure 1. In addition to
the world line of the robot, the figure shows the world line of an 
external object that is the source of its images such
as the stack of cards in Figure 1. 
This source changes its shape at discrete instants
of time demarcated by ticks, running through configurations $\cdots b,
c, d, e, f, g, \cdots$.  
The configuration in each time interval of the object's world line is
labeled to its left. At discrete
instants separated by proper time $\tau_*$, the robot captures an image of
the object. The light rays conveying the image from object to robot are
indicated by dotted lines. The images are stored in the registers $P_0,
P_1, P_2$, and $P_3$ described in Figure 1. The contents of these 
registers in between image captures are displayed in the boxes with $P_0$ on
the left and $P_3$ on the right. 
The history of the contents of the register $P_0$ constitute the
four-dimensional notion of `now' for this robot (the heavily
outlined boxes in the figure). The present is not one
instant along the robot's world line, much less a spacelike surface in
spacetime. Rather, there is a `now' for {\it each} instant along the robot's
world line extending over proper time $\tau_*$. The evolution of the contents
$P_0$ can be described four-dimensionally and is fully consistent with
special relativity. In a similar way the contents of $P_1, P_2$, and $P_3$
constitute a four-dimensional notion of `past'.}
\end{figure}

Figure 3 shows the world line of the robot introduced in Section
\ref{II} 
together with the world line of an object in its environment that appears
in the robot's stored images. The robot illustrated has a short memory with
only four registers $P_0, P_1, P_2, P_3$ whose contents in each interval
$\tau_*$ are indicated by the content of the boxes to
the right of the world line. These contents change at proper time
intervals $\tau_*$ as described in Section \ref{II}. 

The contents of the register $P_0$ defining the robot's present do not
define a spacelike surface representing a moment in time. They do not even
define an instant along the robot's world line since the contents of $P_0$
are consistent over proper time intervals $\tau_*$. Rather, there is content
defining the present for {\it every} instant along the world line. For each
point along the world line the most recently acquired image defines the
present. That is the four-dimensional description of the present.
In a similar way, the data acquired earlier and stored in registers $P_1,
P_2, P_3$ define the robot's past for each point along the world line.

Thus, there is no conflict between the four-dimensional reality of physics
and the subjective past, present, and future of an IGUS. Indeed, as defined
above, the subjective past, present, and future {\it are} four-dimensional
notions. They are not properties of spacetime but of the history 
of a particular IGUS. In
Section \ref{V} we will see that IGUSes constructed differently from our robot
could have different notions of past, present, and future. All of these are
fully consistent with a four-dimensional physical reality.

However, there {\it is} a conflict between ordinary language and the
four-dimensional, IGUS-specific notions of past, present, and future. To
speak of the ``present moment'' of an IGUS, for instance, risks confusion because it
could be construed to refer to a spacelike surface in spacetime stretching 
over the whole universe. No such surface is defined by
physics.\footnote{There is no evidence for preferred frames in spacetime
and modern versions of the Michelson-Morley experiment and other
tests of special relativity set stringent limits on their existence. 
The fractional accuracy of these experiments range down to $10^{-21}$ 
making Lorentz invariance at accessible energy scales one of the
most accurately tested principles in physics. See {\it e.g.} \cite{HW87}.}
  In fact, the ``moment'' in the context of this section
refers to the most recently acquired data of
an IGUS. This is not a notion restricted to one point on the IGUS' world
line which somehow moves along it. Rather is a notion present at {\it every} 
point along the world line.

\subsection{The Common Present}
\label{s3}

The previous discussion has concerned the present, past, and future of
individual localized IGUSes. We now turn to the notion of a common present
that may be held by collections of IGUSes separated in space.

When someone asked Yogi Berra\footnote{Lawrence E.~Berra, a catcher
for the New York Yankees baseball team in the 1950's. See \cite{Ber98}} 
what time it was, he
is reported to have replied ``Do you mean now?'' The laughter usually
evoked by this anecdote shows how strongly 
we hold a common notion of the present. More precisely,
different IGUSes agree on `what is happening now'. This
section is concerned with the limitations on the accuracy of that agreement
arising both from the construction of the IGUSes and the 
limitations of defining simultaneity in special relativity. We continue to
use robot model IGUSes to make the discussion concrete.

\begin{figure}[t]
\centerline{\epsfig{file=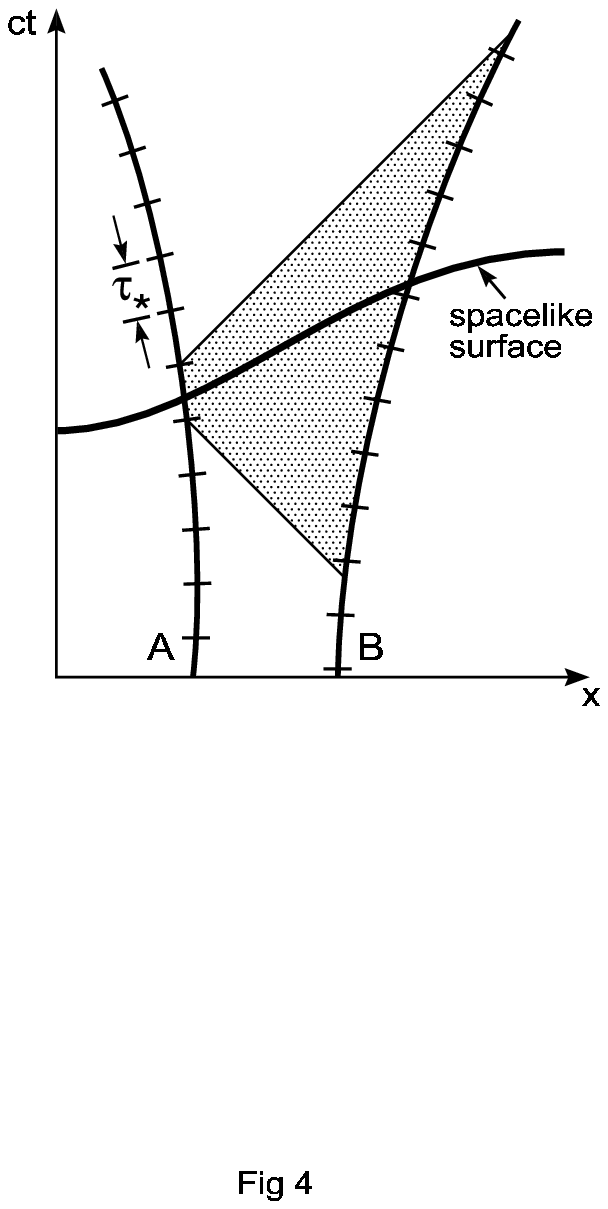,width=3.25in,bbllx=216,bblly=382,bburx=393,bbury=584,clip=true}}
\caption{The ambiguous common present. This spacetime diagram shows the
world lines of two similarly constructed robots $A$ and $B$. The intervals
of proper time of length $\tau_*$ over which the contents of the registers
defining their individual presents are constant are demarcated by ticks. A
common present would be defined by an identification of each interval on
one world line with that on the other `at the same time' (to the accuracy
$\tau_*$). But special relativity allows many different such
identifications. Using the constant $t$ surfaces of the inertial frame
illustrated is one way to define a common present; but any other spacelike
surface such as the one shown would do equally well. The range of ambiguity for
intervals on $B$ that could be said to be `at the same time' as one interval
on $A$ is shown as a shaded region.
The figure shows two robots separated by a distance ({\it e.g.}, as 
defined in the rest
frame of one) over which the light travel time is longer than $\tau_*$
In this situation the ambiguity in the definition of a common present is
much larger than $\tau_*$. However, if the distance between the robots is
much smaller than $c\tau_*$, and if their relative velocity is much less than
$c$ so this continues to be the case, then the ambiguity is much smaller
than $\tau_*$. An {\it approximate} common present can then be defined.}
\end{figure}

Figure 4 shows the world lines of two robots in spacetime together with the
intervals on their world lines that define their individual notions of
`now'. There are at least two reasons that there is no unambiguous notion of a
common `now' that can be shared by the two robots. The first is the
elementary observation\footnote{The ambiguity in the common present
arising from the finite time of IGUS operation has been discussed
in \cite{Cal04} which includes 
a review of current neurophysiological data bearing on this question.} that the present for each individual robot is not
defined to an accuracy better than $\tau_*$. The second reason arises from
special relativity.

A precise common `now' would specify a correspondence between events on the
two world lines. Such a correspondence would specify a notion of
simultaneity between events on the world lines. But there is no unique
notion of simultaneity provided by special relativity. Rather there are
many different notions corresponding 
to the different possible spacelike surfaces that can
intersect the two world lines. Figure 4 illustrates the range of ambiguity.

To illustrate the ambiguity of the present more dramatically, imagine you
are a newscaster on the capital planet Trantor of a galaxy-wide empire some
hundreds of thousands of years in the future. News of events all over the
galaxy pour in constantly via electromagnetic signals. You want to 
broadcast a program called
`The Galaxy Today' reviewing important events in the last 
24 hours (galactic standard). But what time do you assign to the latest news from the
planet Terminus at the edge of the galaxy 60,000 light years away?  There is
an inertial frame\footnote{For example, there is the inertial frame moving
with speed $V$ with respect to the galactic rest frame such that 
\[
\Delta t^\prime = \gamma [\Delta t - V(\Delta x/c)]
\]
where $\gamma = (1-V^2)^{-1/2}$, $\Delta t'=24$hrs, $\Delta t=6 \times 10^4$yr,
and $\Delta x=6\times 10^4$ light years. The required velocity is within a
few parts  in $10^7$ of the velocity of light.} in which those events
happened within the last 24 hours. But in the approximate rest frame of the
galaxy, they happened 60,000 years ago. There is thus no unambiguous notion of
`present' for the collective IGUS consisting of the citizens of the
galactic empire because the ambiguities in defining simultaneity are large
compared to the time scales on which human events happen. You could, of
course, fix on the time of the galactic rest frame as a standard for
simultaneity. But in that case the only comprehensive program you could
broadcast would be `The Galaxy  60,000 Years Ago'. One imagines the
audience for this on Terminus would not be large since the program would 
be seen 120,000 years after the events there happened.

The satellites comprising the Global Positioning System are an example
of a collective IGUS closer to home that faces a similar problem. 
The special relativistic ambiguity in defining
simultaneity for two satellites is of order the light travel time between them
in the approximate inertial frame in which the Earth is at rest. That is
much larger than the light travel time across the few meters accuracy to
which the system aims to locate events. Precise agreement on a definition of
simultaneity is therefore needed. Each satellite clock is corrected so that
it broadcasts the time of a clock on the Earth's geoid (approximately
the ocean surface) \cite{Ash02}.  

No such agreement on a definition of simultaneity appears to be a
prerequisite for the everyday notion of `now' employed by human IGUSes.
Rather, we seem to be employing an approximate, imprecise notion of the
common present appropriate in everyday situations and characterized by the
following contingencies:

\begin{enumerate}

\item The time scale of perception $\tau_*$ is short compared to the time
scales on which interesting features of the environment vary.

\item Individual IGUSes are moving relatively to one another at velocities
small compared to $c$.

\item The light travel time between IGUSes in an inertial frame in which
they are nearly at rest is small compared to the time scales $\tau_*$.

\end{enumerate}
Contingency (1) means that the ambiguity in the `now' of each IGUS is
negligible in the construction of a common present. Contingency (3), based
on (2), means that the ambiguity arising from the definition of
simultaneity is negligible\footnote{These kinds of contingencies and
the synchronization protocols necessary when they are violated have
been discussed in \cite{But84}.}. 

Collections of robots satisfying contingencies 1-3 can agree on what
is happening `now'. Consider just two robots --- Alice and Bob. 
Alice can send Bob a description of the essential features of the
image currently in her register $P_0$. Bob can check whether 
these essential features are the same as those of the image in
his register $P_0$ at the time of receipt. He can then signal back agreement
or disagreement. As long as the light travel time is much shorter than
$\tau_*$ (contingency 3), and as long as the essential features vary 
on much longer time scales (contingency 1) Alice and Bob will agree. 
Contingency 2 ensures that this agreement will persist over an interesting
time scale. Thus Alice and Bob can construct a common present, but it is 
a present that is local, inherently approximate, and contingent upon
their relation to each other and their environment. This approximate 
common `now' is not a surface in spacetime. 

No modification of the laws of physics is needed to understand the common
now of a group of IGUSes as has sometimes been suggested \cite{HH63}. The
common nows of IGUSes meeting the above contingencies will coincide approximately with
constant time surfaces in any inertial frame in which they are
approximately at rest. But these frames are not singled out by the laws of
physics. Indeed, the experimental evidence against preferred frames is special
relativity is extraordinarily good \cite{HW87}. Rather the frames are
singled out by the particular situations of the IGUSes themselves

\section{Why Don't We Remember the Future?}
\label{IV}

The fundamental dynamical laws of physics are invariant under time
reversal to an accuracy
adequate for organizing everyday experience.\footnote{The effective theory
of the weak interactions applicable well below the Planck scale is not 
time-reversal invariant. This is important, for instance, for the synthesis
of baryons in the early universe but negligible for the functioning of our
robot. See, {\it e.g.} \cite{GH93b} for further discussion.} 
They are {\it time neutral}. The Einstein
equation of general relativity and Maxwell's equations for electrodynamics
are examples. But
the boundary conditions specifying solutions to these equations 
describing our universe are not time symmetric. The universe has a smooth
(near homogeneous and isotropic) big bang at one end of time and a very
different condition at the other end. This might be the unending expansion  
driven by a cosmological constant of the simplest cosmological models
favored by observation \cite{Wmap}. Or with different assumptions on
the matter it could be a highly irregular big crunch. In any event, one
end is different from the other\footnote{We thus exclude,
mainly for simplicity, the kind of cosmological model where initial and final
conditions are related by time symmetry that have sometimes been discussed
(e.g \cite{GH93b,Laf93,Cra95,Pri96})}.

%\footnote{Eternal expansion is a clear inference
%from contemporary determinations of the cosmological parameters
%\cite{Wmap} if the dark energy is vacuum energy represented by a 
%positive cosmological constant. The universe could then not be   
%time symmetric about a moment of maximum expansion.}. 
{\it    
By convention, this paper refers throughout to the times closer to the 
big bang as the `past' and times further away as the `future'.} 
 Asymmetry between past and future boundary conditions is
the origin of the various time asymmetries --- `arrows of time' --- 
exhibited by our universe. The arrow of time   
associated with the second law of
thermodynamics is an example\footnote{For reviews of the physics of time
asymmetry from various perspectives, see \cite{GH93b, Dav76, Pen79, Zeh89, 
Pri96}. There is also some discussion in the Appendix. These are only a few
of the references where these issues are treated.} 

The operation of Section \ref{II}'s robot is not time neutral in at least
two respects. First the robot receives information about external events
in its past (closer to the big bang) and not the future\footnote{More 
precisely, registered signals originate from events within the past
light cone of their reception event. We use `future' and `past'
in the present discussion understanding that in each case these
are defined by an appropriate light cone as described in Section
\ref{III}.A.}. 
Second, its processing of the received information is not time
neutral. The flow of information from recording to erasure defines a
direction in time. As mentioned in Section \ref{II}, that direction gives a 
concrete model for the subjective feeling of inexorable forward
progression in time commonly called the `psychological arrow of time'. 
In natural IGUSes, such as ourselves, information flows from past to 
future. More specifically  
the records in registers $P_1, \cdots, P_n$ are of
external events to the {\it past} of those in $P_0$ in decreasing order 
of time from the big bang. This is the reason that Section \ref{II}'s robot
could be said to experience the present, remember the past, and predict
the future\footnote{In Section \ref{II}  we {\it defined} the robot's 
past to be the records in the registers $P_1, \cdots, P_n$. If
information flows from past to future in the robot, that notion 
coincides with the physical past defined as the direction in time 
towards the big bang. From now on we assume this congruence except
where discussing its possible violation, as in this section.} 

Could a robot be constructed that receives information from the future?
Could one be constructed whose psychological arrow of time runs from
future to past with the consequence that it would remember the future?
Both of these possibilities are consistent with time neutral dynamical
laws. But two familiar time asymmetries of our universe prohibit such 
constructions as a practical matter. These are the radiation arrow of
time and the arrow of time associated with the second law of
thermodynamics.

Records of the future {\it are} possible as in a table of future lunar 
eclipses\footnote{We use the term `record' in a
time-neutral sense of an alternative at one time correlated with high
probability with an alternative at another time --- future or past.
Thus, there can be records of the future.}.
Indeed, records of the future are the outcome of any useful process of prediction.
But, our robot's records of the future are obtained by computation, whereas
its records of the past are created by simple, automatic, sensory
mechanisms. These are very different processes both physically and from the
point of view of information processing by the robot.\footnote{Realistic
IGUSes, such as human beings, also create records of the past by computation
as in differing 
interpretations of past experience, and as in the collective construction
of human history, and the history of the universe. However, we did not
explicitly endow our robot with these functions.} By a robot that remembers
the future, we mean one constructed\footnote{In Section \ref{IV}  we will describe
a different construction of a robot which could be said to remember the
future.} as in Section \ref{II} with the records in registers $P_1, \cdots, P_n$
of events to the future of $P_0$.

First consider the question of whether information from the future
could be recorded  by the robot. 
In our universe, electromagnetic radiation is retarded --- propagating
to the future of its emission event.  That time asymmetry is
the radiation arrow of time. The electromagnetic signals
recorded by the robot  propagated to it along the {\it past} light
cone of the reception event.  The images received by the robot, whether of the cosmic background
radiation, distant stars and galaxies, or the happenings in its immediate
environment, are therefore all from past events.
As  far as we know, all other carriers, neutrinos for 
example \cite{Hir87}, are similarly retarded.   
One reason the robot doesn't remember the future is that
it receives no information about it. 

Irrespective of the time its input originates, could a robot 
like that in Section \ref{II} be constructed whose psychological arrow of time 
is reversed, so that internally information flows from future to past? In 
such a robot the events recorded  in registers $P_1, \cdots, P_n$ would 
lie to the
future of that in $P_0$ --- further from the big bang. The robot would 
thus remember the future. Such a construction would run counter to the 
arrow of time
specified by the second law of thermodynamics as we now
review\footnote{Many authors have connected the 
`psychological' arrow of time with that
of the second law.  See, {\it e.g.} \cite{Rei56,Pen79,Haw87}.}.

All isolated subsystems of the universe evolve toward equilibrium. That is
statistics. But the preponderance of isolated systems in our universe are
evolving toward equilibrium from past to future, defining an arrow 
of time.  That is the
second law of thermodynamics that is expressed quantitatively by the 
inexorable increase of an appropriately defined total entropy. 

If the robot processes information irreversibly then its
psychological arrow of time must generally be congruent with the 
thermodynamic arrow of time. The formation of records are crucial steps. 
An increase in total entropy accompanies the
formation of many realistic records. An impact crater in the moon, an
ancient fission track in mica, a darkened photographic grain, or the
absorption of a photon by the retina are all examples. 

But an increase in total entropy is not a necessary consequence of 
forming a record. Entropy increase is necessary only on the erasure of 
a record \cite{Lan61}. 
For the model robot discussed in Section \ref{II}, the only part of
its operation in which entropy must necessarily increase is in the 
erasure of the
contents of the register $P_n$ at each step\footnote{Possibly an
isolated robot could be constructed on the principles of reversible 
computation \cite{Rev} that would not have an erasure step. However,
it seems unlikely that the whole system of robot plus a realistic
observed environment could be reversible.}. However, that is enough. 
To see that, imagine the process of erasure run
backwards from future to past. It would be like bits of smashed
shell reassembling to form an egg. 

To construct a robot with a reversed psychological arrow of time 
it would be necessary to reverse the thermodynamic
arrow,  not only of the robot, but also of the local 
environment it is observing. That is possible in
principle. However, since we have a system of
matter coupled to electromagnetic radiation, it would be necessary to deal
with every molecule and photon within a radius of $2 \times 10^{10}$ km to
reverse the system for a day. More advanced civilizations may find this
amusing. We can have the same fun more cheaply by running the film
through the projector in reverse.
%We should therefore not be surprised that
%naturally-occurring IGUSes are functioning in accord with the arrows of
%time of our universe with the consequence that they remember the past and
%predict the future.\footnote{Other aspects of the subjective difference 
%between past and future can be traced to these arrows of time such as our
%impression that the past is over and done with while the future is subject
%to change.} 

The origin of both the thermodynamic and radiation arrows of time are the
time-asymmetric boundary conditions that single out our universe from the
many allowed by time-reversible dynamical laws. These boundary conditions
connect the two arrows.  A brief sketch of the relevant physics is given in
the Appendix, although it is not necessary for understanding the main
argument of this paper. But it is interesting to think that our subjective
distinction between future and past can ultimately be traced to the
cosmological boundary conditions that distinguish the future and past of
the universe.

\section{Alternatives to  Past, Present, and Future}
\label{V}

The preceding discussion suggests that  the laws of physics
do not define unambiguous notions of past, present, and
future by themselves.  Rather these are features of how specific IGUSes gather and 
utilize information.
%The subjective past, present, and future is consistent with the
% four-dimensional reality of physics and can be described in
%four-dimensional terms.
What then is the origin of the past, present, and future organization of
information in familiar, naturally-occurring IGUSes? Is it the only 
organization compatible with the laws of physics? 
If not, does it arise uniquely from evolutionary
imperatives, or is it a frozen accident that took place in the course of
three billion years of biological evolution? This section discusses such 
questions. 

Certainly some features of the laws of physics are essential prerequisites
to the functioning of Section \ref{II}'s  robot.  There would be no past, present, and
future at all if spacetime did not have timelike directions. 
The fact that IGUSes move on timelike rather than spacelike world lines
is the main part of the reason they can have an approximate common `now'
rather than an approximate common `here'.  An IGUS
functioning in a spacetime where it moved along a closed timelike curve
could not maintain a consistent notion of past and future. 
Likewise, a local distinction between past and future would be difficult to
maintain in the absence of the arrows of time discussed in
Section \ref{II}.
But the features of the physical laws of dynamics and the initial
condition of the universe that are  necessary for a past, present,
future organization of temporal information  are  
consistent with other organizations of this information as
we now show.

\subsection{Different Organizations of Temporal Information}

Perhaps the easiest way of convincing oneself that the notions of past,
present, and future do not follow from the 
laws of physics is to imagine constructing robots that process information
differently from the one described in Section \ref{II}. We consider just
three examples:
\begin{description}

\item {\sl The Split Screen (SS) Robot.} This robot has input to $C$ 
computation from both the most recently acquired data in $P_0$ and
from that in a different register $P_J$ that was acquired a proper 
time $\tau_s \equiv J\tau_*$ earlier along its world line.  There is thus input to conscious 
computation from {\it two} times. 

\item {\sl The Always Behind (AB) Robot.} This robot has input to $C$
computation only from a particular register $P_K, K>0$ and the schema. 
That input is thus always a proper time $\tau_d\equiv K\tau_*$ {\it
behind} the most recently 
acquired data.  

\item {\sl The No Schema (NS) Robot.} This robot has input to $C$
computation from all the registers $P_0, \cdots , P_n$ equally. It
employs no unconscious computation and constructs no schema, but rather takes
decisions by conscious computation from all the data it has.

\end{description}
There seems to be no obstacle to constructing robots wired up in
these ways, and they  process information differently from 
the present, past,  and future organization that we are familiar
with\footnote{Some idea of what the notion of `present' would
be like for some of these robots could be had by serving time in a virtual 
reality suit in which the data displayed was delayed as in the 
$AB$ case, or in which there was an actual split screen as in the
$SS$ case.  An alternative realization of the $SS$ robot's
experience might be produced by  electrical stimulation of the cortex  
that evokes memories of the past which are comparably immediate to 
the present \cite{Penf59}.}. 

An $SS$ robot would have a tripartite division of recorded
information. Its present experience --- its `now' --- would consist of two
times ($P_0$, $P_J$) ---  equally vivid and immediate.   
It would `remember' the intermediate times ($P_1,\cdots,P_{J-1}$), and 
the past ($P_{J+1}, \cdots, P_n $) through the $U$
process of computation and its influence on the schema. 

The $AB$ robot would also have a tripartite division of recorded
information. Its present experience 
would be the contents of the the register $P_K$. It would   
remember the past stored in registers $P_{K+1}, \cdots, P_n$. But
also, it would remember its future stored $P_0, \cdots, P_{K-1}$ a
time $\tau_d$ ahead of its present experience\footnote{That is not in 
conflict with the discussion in Section \ref{IV} because each record is 
still of events in the past to the proper time it was recorded.}. 
(Or perhaps we should say that it would have premonitions of the future.) 

What would discussions with an $AB$ robot be like assuming that our 
information processing is similar to the robot discussed in Section
\ref{II}? Assume for simplicity that we and the $AB$ robot are both 
nearly at rest in one inertial frame and that contingencies 1) through
3) of Section \ref{s3} are satisfied. 
The $AB$ robot would seem a
little slow  --- responding in a time $\tau_d$ or longer to questions. Its
answers to queries about ``What's happening now?'' would seem out
of date. It would be always behind.  

The $NS$ robot would just have one category of recorded information.
Conversations with an $NS$ robot would be impressive since it would
recall every detail it has recorded about the past as immediately and 
vividly as the present\footnote{Perhaps not unlike conversations
with Ireneo Funes in the Borges story {\it Funes the
Memorious} \cite{Bor62}.}. 

The laws of physics supply no obstacle of principle 
to the construction of robots
with exotic organizations of information processing such as the
$SS$, $AB$, and $NS$ robots. But are such organizations a likely outcome of 
biological evolution? Can we expect to find such IGUSes in nature
in this or other planets? We speculate that we will not.  
It is  adaptive for an IGUS of everyday 
size to focus mainly on the most recently acquired data as input to making 
decisions.  The effective low-energy laws of physics in our universe
are {\it local} in spacetime and the nearest data in space and time is
usually the most relevant for what happens next. A frog predicting the
future position of a fly needs the present position and velocity of the
fly, not its location 10s ago. An $AB$ frog would be at a great
competitive disadvantage in not focusing on the most current
information. An $SS$ frog would be wasting precious conscious focus 
on data from the past that is less relevant for immediate prediction
than current data.   

An $NS$ robot is making inefficient use of computing resources in
giving equal focus to present data and data from the past whose
details may not effect relevant future prediction. 
Employing a schema to process the data is plausibly adaptive because 
is a more efficient and faster way of processing data 
with limited computing resources.  The collective
IGUS linked by human culture certainly evolved to make use of schema rather
than focus on the individual records that went into them. For instance,
prediction of the future of the universe is much simpler from a
Friedman-Robertson-Walker model characterized by a few cosmological 
parameters than directly from the records of the measurements that 
determined them.   

\subsection{Different Laws, Different Scales}

Something like the $SS$ organization of temporal information might
be favored by evolution if the laws of physics were not local in
time.  Suppose, for example, that the position of
objects to the future of a  time $t$ depended\footnote{It is not difficult to write down
dynamical difference equations with this property, for instance in 
a one dimensional model we could take
\[
F(t) = m \frac{\left[x(t) - 2x (t-h) + x (t-2h)\right]}{2h^2}.
\]
where $x(t)$ is the body's position, $m$ its mass, and $F(t)$ the
force.  
However, such equations are not consistent with special relativity
and the author is not suggesting a serious
investigation of alternatives to Newtonian mechanics.}, not just on
the force acting and their position and velocity at that time,
but rather on their position at time $t$ and on earlier times 
$t-h$ and $t-2h$ for some fundamental fixed time interval $h$. Then an
organization such as the $SS$ robot with conscious focus on both the  most 
recently acquired data and
that acquired at times $h$ and $2h$ ago might be favored by evolution.

Similarly different organizations might evolve if the IGUS is not
smaller than the scale over which light travels on the
characteristic times of relevant 
change in its environment.  The present, past, future organization
is unlikely to serve such an IGUS well because these notions are
not well defined in these situations, as discussed in Section
\ref{III}.C. As mentioned there, the galactic empires beloved of science 
fiction would be examples of such IGUSes. Faster-than-light travel 
inconsistent with special relativity is often posited by authors
whose stories feature these empires to make their narratives accessible to IGUSes like ourselves 
that do employ --- however, approximately --- a present, past, 
and future organization of information.

\section{Conclusion}
A subjective past, present, and future are not the only
conceivable way an IGUS can organize temporal data in a four-dimensional
physical world consistently with the known laws of physics. But it is a way that may be adaptive for localized IGUSes 
governed by  local physical laws. 
We can conjecture that a subjective past, present, and future is a 
cognitive universal \cite{She94} of such localized IGUSes. That is 
a statement accessible to observational test, at least in principle.

\section{Acknowledgments}
%\acknowledgments

The author thanks Murray Gell-Mann for discussions of complex adaptive
systems over a long period of time. He is grateful to Terry Sejnowski and
Roger Shepard for information about the literature in psychology that bears
on the subject of this paper. Special thanks are due to Roger Shepard
for an extended correspondence on these issues. 
Communications with Jeremy Butterfield, 
Craig Callender and Matt Davidson have been similarly helpful with parts of the philosophical
literature. However, the responsibility rests with the author
for any deficiencies in providing
relevant references to the vast literatures of these subjects of which he
is largely ignorant. Research for this work was supported in part by the 
National Science Foundation under grant PHY02-44764.

\appendix

\section{The Cosmological Origin of Time's Arrows}
\label{app}
 
\renewcommand{\theequation}{\Alph{section}.\arabic{equation}}

The origin of our universe's time asymmetries is not to be found in
the fundamental dynamical laws which are essentially
time-reversible. Rather, both the radiation and the thermodynamic arrows of
time arise from special properties of the initial condition of our
universe\footnote{e.g. Hawking's `no boundary' wave function of the
universe \cite{Haw84}.}. This appendix gives a simplified  discussion 
of these special features 
starting, not at the very beginning, but at the time the hot initial plasma
had become cool enough to be transparent to electromagnetic radiation.
This is the time of ``decoupling'' in cosmological parlance --- about
400,000 years after the big bang or a little over 13 billion years ago.

As Boltzmann put it over a  century ago: ``The second law of thermodynamics
can be proved from the [time-reversible] mechanical theory if one assumes
that the present state of the universe\dots\ started to evolve from an
improbable [{\it i.e.} special] state'' \cite{Bolquote}. The entropy 
of matter and radiation usually defined in
physics and chemistry is about $10^{80}$ 
in the region visible from today at the time
of decoupling (in units of Boltzmann's constant).  This seems high, but it is in fact vastly
smaller than the maximal value of about $10^{120}$ if all that matter
was dumped into a black hole \cite{Pen79}.
The entropy of the matter early in the universe is high because most
constituents are in approximate thermal equilibrium. However, the 
gravitational contribution of the smooth early universe to the entropy
is near minimal, and entropy can grow by the clumping of the matter arising
from gravitational attraction leading to the 
galaxies, stars and other inhomogeneities in the universe we see today. 
 
Amplifying on Boltzmann's statement above, the 
explanation of why
the entropies of isolated subsystems
are mostly increasing in the same direction of time is this: The
progenitors of these isolated systems were all further out of equilibrium at
times closer to the big bang (the past) than they are today. Earlier 
the total entropy was low 
compared to what it could have been.  It has  therefore has tended to 
increase since.  

The radiation arrow of time can also be understood as 
arising from time-asymmetric cosmological boundary conditions applied to
time-reversible dynamical laws. These are Maxwell's equations for 
the electromagnetic field in the presence of charged sources.  
Their time-reversal invariance implies that {\it
any} solution for specified sources at a moment of time can be written at
in {\it either} of two ways: (R) a sum of a free field (no
sources) coming from the past plus {\it retarded} fields whose sources are
charges in the past, or (A) a sum of a free field coming from the future
plus {\it advanced} fields whose sources are charges in the future.
More quantitatively, the four-vector potential
$A_\mu(x)$ at a point $x$ in spacetime can be expressed in the presence of
four-current sources $j_\mu(x)$ in Lorentz gauge as either
\begin{eqnarray*}
A_\mu(x)&=A^{\rm in\ }_\mu (x) + \int d^4 x^\prime D_{\rm ret}
(x-x^\prime)\, j_\mu (x^\prime)\quad &(R) \\
\noalign{\hbox{or}}
A_\mu(x)&=A^{\rm out}_\mu (x) + \int d^4 x^\prime D_{\rm adv}
(x-x^\prime)\, j_\mu (x^\prime) \quad &(A).
\end{eqnarray*}
Here, $D_{\rm ret}$ and $D_{\rm adv}$ are the retarded and advanced Green's
functions for the wave equation and  $A^{\rm in}_\mu (x)$ and $A^{\rm out}_\mu
(x)$ are free fields defined by these decompositions.

Suppose there were no free electromagnetic fields in the distant past 
so that $A^{\rm in\ }_\mu (x) \approx 0$.
Using the R description above, this time asymmetric boundary condition
would imply that present fields can be entirely ascribed to sources in the
past. This is retardation and that is the electromagnetic arrow of time.

This explanation needs to be refined for our universe because, at least if
we start at decoupling, there {\it is} a significant amount of free 
electromagnetic
radiation in the early universe constituting the cosmic background
radiation (CMB). Indeed, at the time of decoupling the energy density in
this radiation was approximately equal to that of matter. Even today,
approximately 13
billion years later, after being cooled and diluted by the expansion
of the universe, the CMB is still the
largest contributor to the electromagnetic energy density in the universe
by far.

The CMB's spectrum is very well fit by a black body law 
\cite{Fix96}. That strongly suggests that the radiation is
{\it disordered} with maximal entropy for its energy density. 
There is no evidence for the kind of correlations (sometimes called
`conspiracies') that would tend to cancel $A^{\rm out}_\mu (x)$ in 
the far future and give rise to advanced rather than retarded
effects\footnote{For an experiment that checked on advanced 
effects see \cite{Par73}.}.    

The expansion of the universe has red-shifted the peak luminosity of the 
CMB at decoupling to microwave wavelengths today. There is thus a 
negligible amount energy left over from the big bang in the wavelengths 
we use for vision, for instance. The radiation used by realistic 
IGUSes is therefore retarded. A contemporary robot functioning
at wavelengths where the CMB is absent will therefore be receiving 
information about charges in the past.  This selection of 
wavelengths is plausibly not accidental but adaptive. A   
contemporary robot seeking to function with input from microwave
wavelengths would find little emission of interest, and what there
was would be overwhelmed by the all-pervasive CMB, nearly equally bright in
all directions, and carrying no information.

\end{document}